# Coulomb Blockade from the Shell of an InP-InAs Core-Shell Nanowire with a Triangular Cross Section

D. J. O. Göransson[1], M. Heurlin[1], B. Dalelkhan[1], S. Abay[1], M. E. Messing[1], V.F. Maisi[1], M. T. Borgström[1], H. Q. Xu[1,2,3,*]

[1] *NanoLund, Division of Solid State Physics, Lund University, Box 118, S-22100 Lund, Sweden*
[2] *Beijing Key Laboratory of Quantum Devices, Key Laboratory for the Physics and Chemistry of Nanodevices, and Department of Electronics, Peking University, Beijing 100871, China*
[3] *Beijing Academy of Quantum Information Sciences, West Bld. #3, No.10 Xibeiwang East Rd., Haidian District, Beijing 100193, China*

[*]Corresponding author. Email: hqxu@pku.edu.cn or hongqi.xu@ftf.lth.se

(Date: January 16, 2019)

**ABSTRACT**

We report on growth of InP-InAs core-shell nanowires and demonstration of the formation of single quantum structures, which show Coulomb blockade effect, over entire lengths of the nanowires. The core-shell nanowires are grown by a selective area growth technique via metal-organic vapor phase epitaxy. The as-grown core-shell nanowires are found to be of wurtzite crystals. The InP cores have a hexagonal cross section, while the InAs shell are grown preferentially on specific $\{1\bar{1}00\}$ facets, leading to the formation of the core-shell nanowires with an overall triangular cross section. The grown core-shell nanowires are transferred on to a Si/SiO$_2$ substrate and then contacted with several narrow metal electrodes. Low-temperature transport measurements show the Coulomb-blockade effect. We analyze the measured gate capacitance and single electron charging energy of the devices and demonstrate that a quantum structure which shows the Coulomb blockade effect of a many-electron quantum dot is formed over the full length of a single core-shell nanowire and consists of the entire InAs shell in the nanowire.

Semiconductor nanowires (NWs) have been extensively investigated for potential applications in optoelectronics, nanoelectronics, and quantum devices.[1–7] An intriguing advantage of the semiconductor NW technology is the presence of an ultralow density of dislocations in NW heterostructures, enabling building NW heterostrutures from materials with a large difference in lattice constant and accommodating large built-in strains in them.[8,9] It is also possible to achieve complex NW architectures by growing selectively materials on different facets.[10] It has been shown that heterostructured core-shell NWs with a built-in strain can possess an increased carrier mobility[4,11] or a piezoelectric field along the NW axis.[12,13] The former is desired for high-speed electronic applications and the latter has a prospective application in novel photovoltaic technology. Heterostructured core-shell NWs are also predicted to yield favorable designs of quantum systems in which Majorana bound states[14] and Aharonov-Bohm oscillations[15,16] can be detected. Recent low-temperature transport measurements have demonstrated the observations of the Aharonov-Bohm states in GaAs-InAs core-shell NWs.[17,18] Low-temperature transport measurements have also demonstrated the observation of induced superconductivity in superconductor-contacted Ge-Si[19] and GaAs-InAs core-shell NWs.[20]

Metal-organic vapor phase epitaxy (MOVPE) is one of the most common techniques to grow NWs with excellent controls of crystal structures and NW morphology. Two NW growth mechanisms employed within MOVPE are metal particle assisted vapor-liquid-solid (VLS) growth and selective area epitaxy (SAE) growth.[21] The majority of NW growth research has focused on particle assisted VLS growth and the technique has been used to grow NW p-n junctions[7,22] and heterostructures.[23,24,22] However, in SAE, NW growth can be achieved without risk of metal contamination, due to the absence of metal seed particles, and shell overgrowth in achieving core-shell NWs can be made with relatively less technical difficulty. Furthermore, InP NWs grown by SAE have been shown to have low surface recombination velocity and to be of stacking fault free crystals with excellent optical quality and high lasing efficiency.[25] It has also been demonstrated that with SAE, it is possible to achieve no-tapered, pure crystal phase InP NWs.[26]

In this letter, we report on growth of high-quality InP-InAs core-shell NWs with a triangular cross section via MOVPE and formation of good-quality quantum shell structures over entire lengths of the NWs. The core-shell NWs are grown by SAE technique and are of wurtzite crystal structure. Devices for transport measurements are fabricated by transferring as-grown core-shell NWs on to a Si/SiO$_2$ substrate and then contacting them with several Ti/Au metal stripe electrodes on each. The transport measurements are performed at a low temperature in a dilution refrigerator. The measurements show the Coulomb blockade effect with the charge stability diagrams characterized by a series of well-defined, equally sized Coulomb diamonds in each device—a characteristic often observed in a many-electron quantum dot. We also measure the charge stability diagrams using different configurations of source and drain contact electrodes and extract device characteristic parameters, such as gate capacitance and charging energy. The results show that the Coulomb blockade seen in each device arises from single electron charging to the entire InAs shell of the NW, i.e., a quantum structure that shows the Coulomb blockade effect of a quantum dot is formed over the entire InAs shell in each NW.

Our InP-InAs core-shell NW growth experiment starts with covering an InP (111)A single crystal wafer, i.e., a growth substrate, with a 27-nm-thick layer of $SiN_x$ via plasma enhanced chemical vapor deposition. After subsequently spin coating of 13-nm-thick resist LOR A and 200-nm-thick resist CSAR 62, circular holes are patterned by electron beam lithography (EBL). Reactive-ion etching (RIE) is used to etch down the $SiN_x$ layer with the use of the EBL patterned double resist layer as an etch mask. The resists are then removed in 1-methyl-2-pyrrolidinone based solvent (Remover 1165) at 90 °C and subsequently the growth substrate is cleaned by rinses in IPA and deionized water, and by oxygen plasma ashing. The preparation of the growth substrate is finalized by polishing the patterned $SiN_x$ mask in diluted HF (diluted in water, 1:100) for 30 s. The NW growth by SAE in the mask opening areas is performed in a horizontal flow MOVPE reactor operated at 100 mBar with a total gas flow rate of 6 l/min. Trimethylindium (TMIn), phosphine and arsine are used as source gases. First, the substrate is annealed at 750 °C for 10 min under phosphine flow. The InP NW cores are then grown for 10 min at 710 °C with the molar fractions of TMIn and phosphine set to $1.2 \times 10^{-5}$ and $2.5 \times 10^{-2}$, respectively. Then the group III precursor supply is switched off and the temperature is lowered to 460 °C at which phosphine is switched off and arsine is switched on. Then, TMIn is reintroduced into the reactor to grow InAs shells for 5 min under the molar fractions of TMIn and arsine set to $6.7 \times 10^{-6}$ and $8.4 \times 10^{-4}$, respectively. The sample is then cooled down in the reactor under arsine flow in order to prevent the InAs surfaces from decomposition.

The morphology and crystal structure of as-grown core-shell NWs are investigated using a scanning electron microscope (SEM) and a transmission electron microscope (TEM). Figure 1(a) shows an SEM image (top view) of as-grown InP NW cores without overgrowth of InAs shells. Here, the solid arrows represent the crystallographic directions of the InP substrate and the orientations of an InP NW facet, and the dashed line indicates the orientation of a cleaved edge of the substrate. It is seen that the InP cores are grown vertically on the substrate and are arranged in a square lattice. It is also seen that the cores have a hexagonal cross section (for a clearer view, see a zoom-in SEM image and a schematic given in the upper-right inset). The side facets of the InP cores are seen to be rotated 30º with respect to the cleaved edge of the substrate which is assumed to be of cubic {110} types [see the dashed line in Fig. 1(a)], showing that the InP NW side facets are of hexagonal {1$\bar{1}$00} planes. Figure 1(b) shows a top-view SEM image of as-grown InP-InAs core-shell NWs and Fig. 1(c) is a 30º-tilted-view SEM image of as-grown core-shell NWs. Here, in Fig. 1(b), a few crystallographic directions of the InP substrate, the orientation of a core-shell NW facet, and the orientation of a substrate cleaved edge are again indicated. It is seen in both Fig. 1(b) and Fig. 1(c) that for most core-shell NWs, the shell in each individual NW grows selectively on three of the six InP core NW facets much faster than on other three facets. Thus, an entire core-shell NW has a triangular shaped cross section with three triangular pockets of InAs located in its three corners [see the schematic view in the upper-right inset in Fig. 1(b)]. As a result, it is most likely that the shell in such a core-shell NW forms a coupled triple InAs NW with the couplings achieved through three thin layers of InAs. The triangularly shaped core-shell NW have a size of $D$ = 200-225 nm, as determined by SEM measurements [cf. Fig. 1(b)]. The hexagonal cross-sectional cores have a diameter of $d$ = 110-120 nm, determined from the reference growth samples without overgrowth shells of InAs. It should be noted that there exists a small fraction of core-shell wires with an overall hexagonal cross section on the growth substrate. A mixture of hexagonally and triangularly shaped core-shell wires has also been found in growth of InP-InGaAs core-shell NWs.[27]

Figure 1(d) shows a representative high-resolution TEM image of a top segment of an as-grown NW. Here, it is seen that the NW is a high-quality single crystal with only a few stacking faults formed locally in the top region. Figures 1(e) and 1(f) show the results of the fast Fourier transform analyses made on a region sufficiently far from the top end marked by square 1 in Fig. 1(d) and in the top end region marked by square 2 in Fig. 1(d), respectively. Note that the two regions are separated by the localized stacking faults. It is shown that a small cap of InAs with a zincblende crystal structure is formed at the top of the NW. Away from the top, a high-quality InP-InAs core-shell NW structure is formed and is in pure wurtzite crystalline phase. The formation of the InP-InAs core-shell NWs structure has also been confirmed by energy-dispersive X-ray spectroscopy (EDX) measurements (not shown here). In addition, by EDX line scans across as-grown NWs, the InP core size is determined to be $d = 110 - 120$ nm, in good agreement with the results obtained by the SEM measurements of the reference InP NW samples.

For device fabrication, as-grown InP-InAs core-shell NWs are transferred on to an *n*-type Si substrate (with a resistivity in a range of 1-5 × $10^{-3}$ Ω cm) covered by a 130-nm-thick layer of thermal $SiO_2$ on top. The backside of the substrate is etched using buffered oxide etchant (BOE) to remove the oxide, and is then coated with Au to form a contact to the Si, which is used as a global back gate. After NW transferring, SEM images are acquired for locating NWs with high accuracy against predefined markers on the substrate. Then, EBL is used to define the patterns of the contracts to the NWs selected for the device fabrication. Here, it is important to note that the triangular cross-sectional shape of the NWs allows for placing contacts on sides of the NWs, enabling achieving several different source-drain contact configurations on each NW. The sample is subsequently cleaned by oxygen plasma etching to remove EBL resist residues and by etching in a water mixture of ammonium sulfide $(NH_4S_x)$[28] to remove a thin layer of native surface oxide in the contact areas of the NWs. Here, three-molar Sulphur is added to a 20% $NH_4S_x$ stock solution. The solutions is then diluted in water in the ratio of 1:79, a few minutes prior to etching, and then heated to 42 °C, and the etching time used is 5 min. After etching, the sample is rinsed in deionized water and is then immediately loaded into a thermal evaporator. The contact metal layers of 10-nm-thick Ti and 120-nm-thick Au are evaporated in the thermal evaporator. The device fabrication is completed by lift-off in Remover 1165 at 70 °C for 30 min. Figures 2(a)-2(c) show a fabricated device and the device schematics in cross-sectional side view and top view. This device has four contacts placed on two sides of the NW in a rectangular configuration [see Figs. 2(a) and 2(c)], enabling transport measurements both across the NW and along the NW axis.

The fabricated devices are characterized at room temperature to extract the NW resistance and the contact resistance. In the present work, the devices with the room-temperature NW (unit length) resistance of 0.8-1.2 kΩ/µm and the room-temperature contact-pair resistance of 3.2-3.8 kΩ are selected for low temperature measurements in a dilution refrigerator. This range of contact resistance is chosen in order to obtain a large low-temperature contact resistance (higher than the quantum resistance of ~13 kΩ), as is normally required for observation of the Coulomb blockade effect. Several devices have been measured at low temperature in the dilution refrigerator and below we will only present the results of measurements for the device shown in Fig. 2(a).

Figure 2(d) displays the measured source-drain current $I_{sd}$ of the device shown in Fig. 2(a) as a function of back-gate voltage $V_{bg}$ at a source-drain voltage of $V_{sd} = 22$ µV applied between contacts E and F and at a base temperature of $T = 14$ mK. The other two contacts, i.e., contacts A and B, are left floating during the measurements. The measurements show regular current oscillations with a series of equally spaced sharp current peaks and well-defined wide low current valleys, i.e., the typical Coulomb blockade effect of a quantum dot in the many-electron regime.

To further identify the origin of the Coulomb blockade effect, we measure the differential conductance $dI_{sd}/dV_{sd}$ of the device as a function of source-drain voltage $V_{sd}$ and back-gate voltage $V_{bg}$ (charge stability diagram) at $T = 14$ mK and at different source-drain contact configurations. The results of the measurements are shown in Fig. 3. Figure 3(a) shows the measured charge stability diagram of the device with the source-drain voltage applied to contacts E and F (i.e., along the NW axis) while keeping the remaining contacts A and B floating. Here, the measured charge stability diagram is represented by a series of almost identical Coulomb diamond structures with low differential conductance values inside the diamonds, indicating again that a quantum structure that shows a many-electron quantum dot behavior is formed in the NW. Note that increased differential conductance seen in the upper and lower corners of each diamond results from strong co-tunneling effect due to good transparency of the contacts. Note also that an abrupt change in the differential conductance is seen on the left side of the figure. This is due to a sudden charge arrangement in the gate dielectric and such a charge arrangement is commonly seen in a gate defined or a gate tuned quantum dot device [see also our measurements shown in Figs. 3(b) and 3(c)]. Figure 3(b) shows the measured charge stability diagram with the source-drain bias voltage applied along the NW axis but to contacts A and B (i.e., the electrodes fabricated on the other side of the NW) while keeping contacts E and F floating. Figures 3(c) and 3(d) show the charge stability diagram measurements with the source-drain bias voltage applied across the NW, i.e., to contacts A and E while keeping contacts B and F floating, and to contacts B and F while keeping contacts A and E floating, respectively. All these measurements show almost the same Coulomb diamond structures as observed in Fig. 3(a). For example, the size of the Coulomb diamonds in back-gate voltage, $\Delta V_{bg} = 2.0$ mV, is the same for the measurements with all the four contact configurations, i.e., with the bias voltage applied both along and across the NW. As the gate-voltage spacing $\Delta V_{bg}$ is independent of the source-drain bias voltage configuration, we can conclude that the same quantum dot structure with its extension over the entire length of the NW is measured in Figs. 3(a) to 3(d). Because of the large conduction band offset between the InP and InAs materials, the quantum dot structure can be identified as the entire InAs shell in the core-shell NW. In the following we will sometimes interchangeably use both the quantum dot and the entire InAs shell of the NW to refer to the quantum object which give the Coulomb blockade effect observed in this work. The capacitance of the back-gate coupling to the quantum structure is extracted, from the measured back-gate voltage spacing of the Coulomb diamonds, to be $C_g = e/\Delta V_g = 80$ aF. From the size of the Coulomb diamonds in source-drain bias voltage, the single electron charging energy of the quantum structuret (which is almost the same as the electron addition energy in this many electron regime) is extracted as $E_c \approx 250$ μeV.

The back-gate capacitance to the quantum structure could also be estimated based on an analytical model given in Ref. 29. This analytical model has been commonly applied in studies of back-gated NW field effect transistors.[30] Numerical calculations performed for back-gated triangular NWs in Ref. 31 could also be used to estimate the back-gate capacitance to the quantum structure and to verify that the quantum structure is formed over the full length of the NW. According to the model of Ref. 29, the capacitance of the back gate, which is assumed to be an infinite plate, to the NW is given by

$$C_g = 2\pi L \varepsilon_0 \varepsilon_e / \operatorname{acosh}\left(\frac{t_{ox}+r}{r}\right), \quad (1)$$

where $L$ is the length of the NW, $\varepsilon_0$ is the vacuum permittivity and $\varepsilon_e$ is the effective relative permittivity of the oxide and vacuum surrounding the NW, $r$ is the NW radius and $t_{ox}$ is the gate oxide thickness. Without using an effective dielectric constant $\varepsilon_e$ the model yields an overestimation of the capacitance with approximately a factor of two for a cylindrical NW, since the NW is surrounded by the oxide only from one side.[31] Assuming an effective radius of $r = 78$ nm, in order to have an equal cross-sectional area as the equilateral triangle with a size of 210 nm, and $\varepsilon_e = 1.95$,[31] we obtained, based on the analytical model of Eq. (1), $C_g/L = 66$ aF/μm. By extrapolating the numerical data in Ref. 31 to the NW size/oxide thickness ratio of $D/t_{ox} = 1.6$, we obtain a value of $C_g/L = 69$ aF/μm, which is very close to the value obtained based on the analytical model. From the experimental data shown in Fig. 3, we determine a value of $C_g/L = 59$ aF/μm when the full length of the NW, $L=1.357$ μm, is used. This value is in good agreement with the values obtained above based on the analytical model and the numerical results. Such a good agreement provides a strong support to our conclusion that the quantum structure which shows the Coulomb blockade effect of a quantum dot is formed over the full length of the NW. Finally, we would like to note that the charge stability diagram has been measured for the device over a large range of back-gate voltages and similar Coulomb diamond structures as seen in Figure 3 are observed.

In conclusion, high quality, stacking fault free wurtzite crystalline InP-InAs core-shell NWs have been grown by SAE technique via MOVPE. The InP cores have a hexagonal cross section, while the InAs shells are grown on different facets of the InP cores with different growth rates, leading to the formation of the core-shell NWs with a triangular cross section. The NWs are transferred on to a Si/SiO$_2$ substrate and are then contacted by narrow metal electrodes on the side surfaces. The low temperature transport measurements of the fabricated devices show the Coulomb blockade effect with their charge stability diagrams characterized by a series of well-defined, equally sized Coulomb diamond structures. We have also carried out the charge stability diagram measurements using different combinations of the narrow local contacts and analyzed the gate capacitance and single electron charging energy. The results show that a quantum structure which shows the Coulomb blockade effect of a many-electron quantum dot is formed over the entire length of a single core-shell NW and consists of the entire InAs shell. A promising future direction would be to grow an InAs shell with controlled thicknesses on different core facets, allowing to adjust the lateral coupling between three InAs quantum wires formed in the corners of the triangle from a weak non-coupled case to a strong coupling case, and to study proximity-induced topological superconducting states in such a core-shell NW.

The authors would like to thank Dr. Adam Burke and Dr. Claes Thelander for their valuable assistances with the low temperature measurement setups. This work was performed within NanoLund with support from Myfab and was financially supported by the Swedish Research Council (VR), the Ministry of Science and Technology of China through the National Key Research and Development Program of China (Grant Nos. 2017YFA0303304 and 2016YFA0300601), and the National Natural Science Foundation of China (Grant Nos. 11874071, 91221202, and 91421303).

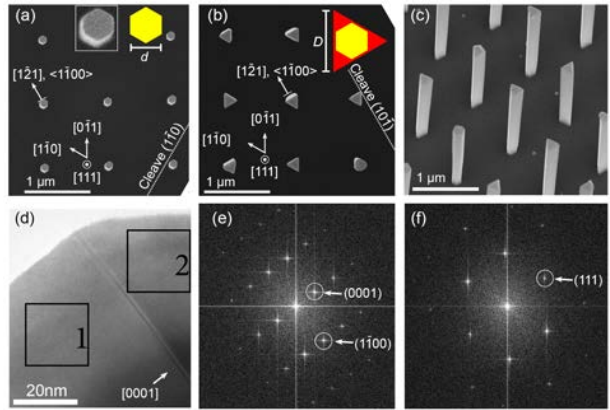

Fig 1. (a), SEM image (top view) of as-grown InP cores without the shell overgrowth. These InP core NWs have a hexagonal cross section and a size of $d$ = 110 - 120 nm. The upper-right inset shows a zoom-in top-view image of an InP core NW and its schematic. The solid arrows mark the crystallographic directions of the InP substrate and the orientations of a hexagonal InP NW facet, and the dashed line marks the orientation of a cleaved edge of the substrate which is of a {110} plane. (b) SEM image (top view) of as-grown InP-InAs core-shell NWs. The upper-right inset shows a schematic view of an as-grown InP-InAs core-shell NW with shells of InAs marked in red. The core-shell NWs have a triangular cross section and the shells of InAs are grown as coupled triple NWs. Here, a few crystallographic directions of the InP substrate, the orientation of a core-shell NW facet, and the orientation of a substrate cleaved edge are again indicated. (c) SEM image (30° tilted view) of the InP-InAs core-shell NWs. (d) High resolution TEM image of a top segment of an InP-InAs core-shell NW. (e) and (f) Fast Fourier transform spectroscopies taken in the region marked by square 1 and in the region marked by square 2 in (d). The core-shell segment of the NWs are wurtzite crystals and are grown along the [0001] direction as shown in (e). The top part is a zincblende InAs crystal as identified in the fast Fourier transform shown in (f). A few localized stacking faults are found which separate pure wurtzite core-shell NWs and top zincblende InAs caps.

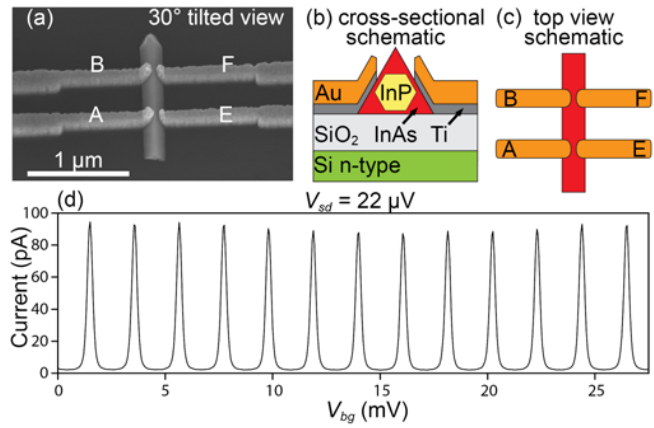

Fig 2. (a) SEM image (30° tilted view) of the InP-InAs core-shell NW device measured at low temperature for this work. (b) Schematic cross-sectional view of the NW device with the contact structure displayed. (c) Schematic top-view of the NW device with the Ti/Au contact arrangement displayed. (d) Source-drain current measured as a function of back-gate voltage $V_{bg}$ with a source-drain bias voltage of $V_{sd} = 22\ \mu V$ applied to contacts E and F at base temperature $T = 14$ mK. Contacts A and B are left floating in the measurements.

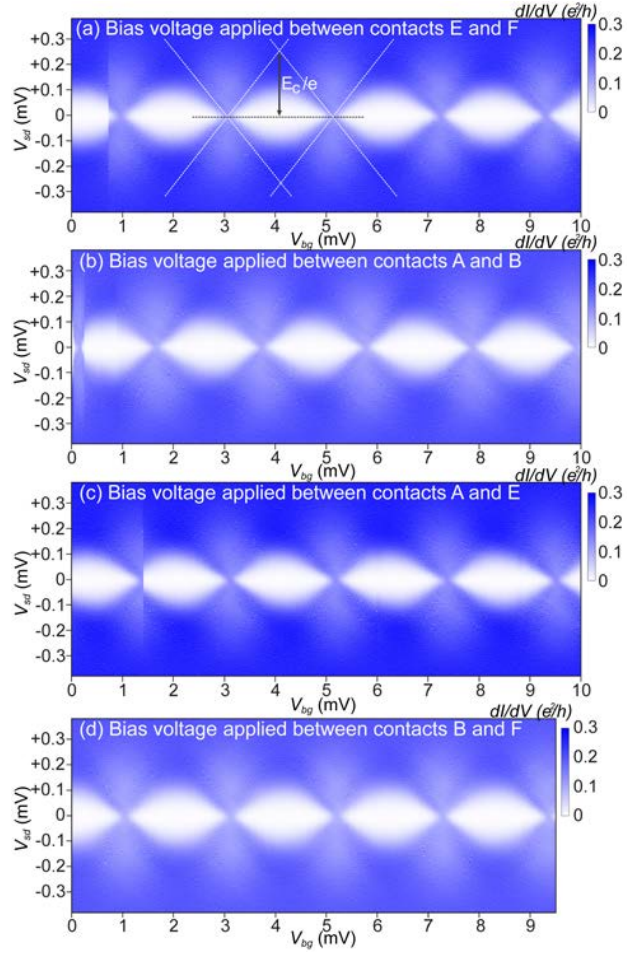

Fig. 3. Differential conductance $dI_{sd}/dV_{sd}$ (on color scale) of the device shown in Fig. 2(a) measured as a function of back-gate voltage $V_{bg}$ and source-drain voltage $V_{sd}$ (charge stability diagram) at $T$=14 mK, back-gate voltages around $V_{bg} = 0$, and source-drain voltages applied (a) using contacts E and F while keeping contacts A and B floating, (b) using contacts A and B while keeping contacts E and F floating, (c) using contact A and E while keeping contact B and F floating, and (d) using contacts B and F while keeping contacts A and E floating. Similar Coulomb diamond structures of the differential conductance are seen in all the measured charge stability diagrams. Abrupt changes in the differential conductance seen in (a), (b) and (c) are due to sudden charge arrangements in the gate dielectric and such a charge arrangement is commonly seen in a gate defined or a gate tuned quantum dot device.